\newif\ifJOURNAL
\JOURNALfalse
\newif\ifarXiv
\arXivfalse
\newif\ifWP
\WPfalse
\newif\ifFULL
\FULLfalse
\newif\ifLATIN
\LATINfalse

\arXivtrue

\ifarXiv\LATINtrue\fi	

\newif\ifnotJOURNAL	
\notJOURNALtrue
\ifJOURNAL\notJOURNALfalse\fi

\newif\ifnotarXiv	
\notarXivtrue
\ifarXiv\notarXivfalse\fi

\newif\ifTR		
\TRfalse
\ifarXiv\TRtrue\fi
\ifWP\TRtrue\fi
\newif\ifnotTR
\notTRtrue
\ifarXiv\notTRfalse\fi
\ifWP\notTRfalse\fi

\newif\ifnotFULL	
\notFULLtrue
\ifFULL\notFULLfalse\fi

\newif\ifnotLATIN	
\notLATINtrue
\ifLATIN\notLATINfalse\fi

\ifJOURNAL
  \newcommand{\CTI}{GTP24}
\fi
\ifarXiv
  \newcommand{\CTI}{vovk:arXiv0712.1275}
\fi
\ifWP
  \newcommand{\CTI}{GTP24}
\fi

\ifnotLATIN
  \newcommand{\Takeuchi}{takeuchi:2004}
\fi
\ifLATIN
  \newcommand{\Takeuchi}{takeuchi:2004latin}
\fi

\ifJOURNAL
\documentclass{article}
\usepackage{amsmath,amsthm,amsfonts,amssymb,latexsym}
\newcommand{\Extra}[1]{}
\fi

\ifarXiv
\documentclass{article}
\usepackage{amsmath,amsthm,amsfonts,amssymb,latexsym}
\newcommand{\Extra}[1]{}
\fi

\ifWP
\documentclass[toc]{gtarticle}
\usepackage{amsmath,amsthm,amsfonts,amssymb,latexsym,epsfig}
\renewcommand{\Extra}[1]{}
\fi

\ifFULL
\usepackage{color}
\renewcommand{\Extra}[1]{\blue{#1}}

\newcommand{\blue}[1]{\textcolor{blue}{#1}}
\newcommand{\bluebegin}{\begingroup\color{blue}}
\newcommand{\blueend}{\endgroup}

\fi

\allowdisplaybreaks[4]

\newcommand{\Vladimir}{Vladimir}

\ifnotLATIN
  \input{OT2enc.def}
  
  \usepackage{CJK}
\fi

\newcommand{\st}{\mathrel{|}}		

\renewcommand{\And}{\mathrel{\&}}	

\newcommand{\dd}{\mathrm{d}}		

\newcommand{\K}{\mathcal{K}}		

\newcommand{\FFF}{\mathcal{F}}		

\DeclareMathOperator{\III}{\mathbb{I}}		

\newcommand{\bbbp}{\mathbb{P}}		

\DeclareMathOperator{\UpProb}{\overline{\bbbp}}		

\DeclareMathOperator{\var}{var}		
\DeclareMathOperator{\vex}{vex}		
\DeclareMathOperator{\nc}{nc}		

\newcommand{\bbbn}{\mathbb{N}}		
\newcommand{\bbbr}{\mathbb{R}}		
\newcommand{\bbbz}{\mathbb{Z}}		

\theoremstyle{plain}
\newtheorem{theorem}{Theorem}

\newtheorem{lemma}{Lemma}

\theoremstyle{definition}

\title{Continuous-time trading and\\emergence of volatility}

\ifarXiv
\author{Vladimir Vovk\\
\texttt{vovk{\rm@}cs.rhul.ac.uk}\\
\texttt{http://vovk.net}}
\fi

\ifWP
\author{Vladimir Vovk}

\twodatestrue

\fi

\begin{document}

\maketitle

\begin{abstract}
  This note continues investigation of randomness-type properties
  emerging in idealized financial markets with continuous price processes.
  It is shown,
  without making any probabilistic assumptions,
  that the strong variation exponent of non-constant price processes
  has to be 2,
  as in the case of continuous martingales.
\end{abstract}

\section{Introduction}
\label{sec:introduction}

This note is part of the recent revival of interest in game-theoretic probability
(see, e.g., \cite{shafer/vovk:2001,\Takeuchi,kumon/etal:2007,horikoshi/takemura:2008,kumon/takemura:2008}).
It concentrates on the study of the ``$\sqrt{\dd t}$ effect'',
the fact that a typical change in the value of a non-degenerate diffusion process
over short time period $\dd t$ has order of magnitude $\sqrt{\dd t}$.
Within the ``standard'' (not using non-standard analysis) framework
of game-theoretic probability,
this study was initiated in \cite{takeuchi/etal:2007}.
In our definitions, however,
we will be following \cite{\CTI},
which also establishes some other randomness-type properties
of continuous price processes.
The words such as ``positive'', ``negative'', ``before'', and ``after''
will be understood in the wide sense of $\ge$ or $\le$,
respectively;
when necessary, we will add the qualifier ``strictly''.

\ifarXiv
  The latest version of this working paper can be downloaded from the web site
  \texttt{http://probabilityandfinance.com}
  (Working Paper 25).
\fi

\section{Null and almost sure events}
\label{sec:definitions}

We consider a perfect-information game between two players,
Reality (a financial market) and Sceptic (a speculator),
acting over the time interval $[0,T]$,
where $T$ is a positive constant fixed throughout.
First Sceptic chooses his trading strategy
and then Reality chooses a continuous function $\omega:[0,T]\to\bbbr$
(the price process of a security).

Let $\Omega$ be the set of all continuous functions $\omega:[0,T]\to\bbbr$.
For each $t\in[0,T]$,
$\FFF_t$ is defined to be the smallest $\sigma$-algebra
that makes all functions
$\omega\mapsto\omega(s)$, $s\in[0,t]$, measurable.
A \emph{process} $S$ is a family of functions
$S_t:\Omega\to[-\infty,\infty]$, $t\in[0,T]$,
each $S_t$ being $\FFF_t$-measurable
(we drop the adjective ``adapted'').
An \emph{event} is an element of the $\sigma$-algebra
$\FFF_T$.
Stopping times $\tau:\Omega\to[0,T]\cup\{\infty\}$
w.r.\ to the filtration $(\FFF_t)$
and the corresponding $\sigma$-algebras $\FFF_{\tau}$
are defined as usual;
$\omega(\tau(\omega))$ and $S_{\tau(\omega)}(\omega)$
will be simplified to $\omega(\tau)$ and $S_{\tau}(\omega)$,
respectively
(occasionally,
the argument $\omega$ will be omitted
in other cases as well).

The class of allowed strategies for Sceptic is defined in two steps.
An \emph{elementary trading strategy} $G$
consists of an increasing sequence of stopping times
$\tau_1\le\tau_2\le\cdots$
and, for each $n=1,2,\ldots$,
a bounded $\FFF_{\tau_{n}}$-measurable function $h_n$.
It is required that, for any $\omega\in\Omega$,
only finitely many of $\tau_n(\omega)$ should be finite.
To such $G$ and an \emph{initial capital} $c\in\bbbr$
corresponds the \emph{elementary capital process}
\begin{equation*}
  \K^{G,c}_t(\omega)
  :=
  c
  +
  \sum_{n=1}^{\infty}
  h_n(\omega)
  \bigl(
    \omega(\tau_{n+1}\wedge t)-\omega(\tau_n\wedge t)
  \bigr),
  \quad
  t\in[0,T]
\end{equation*}
(with the zero terms in the sum ignored);
the value $h_n(\omega)$ will be called the \emph{portfolio}
chosen at time $\tau_n$,
and $\K^{G,c}_t(\omega)$ will sometimes be referred to
as Sceptic's capital at time $t$.

A \emph{positive capital process} is any process $S$
that can be represented in the form
\begin{equation}\label{eq:positive-capital}
  S_t(\omega)
  :=
  \sum_{n=1}^{\infty}
  \K^{G_n,c_n}_t(\omega),
\end{equation}
where the elementary capital processes $\K^{G_n,c_n}_t(\omega)$
are required to be positive, for all $t$ and $\omega$,
and the positive series $\sum_{n=1}^{\infty}c_n$ is required to converge.
The sum (\ref{eq:positive-capital}) is always positive
but allowed to take value $\infty$.
Since $\K^{G_n,c_n}_0(\omega)=c_n$ does not depend on $\omega$,
$S_0(\omega)$ also does not depend on $\omega$
and will sometimes be abbreviated to $S_0$.

The \emph{upper probability} of a set $E\subseteq\Omega$
is defined as
\begin{equation*}
  \UpProb(E)
  :=
  \inf
  \bigl\{
    S_0
    \bigm|
    \forall\omega\in\Omega:
    S_T(\omega)
    \ge
    \III_E(\omega)
  \bigr\},
\end{equation*}
where $S$ ranges over the positive capital processes
and $\III_E$ stands for the indicator of $E$.

We say that $E\subseteq\Omega$ is \emph{null} if $\UpProb(E)=0$.
A property of $\omega\in\Omega$ will be said to hold \emph{almost surely} (a.s.),
or for \emph{almost all $\omega$},
if the set of $\omega$ where it fails is null.

Upper probability is countably (and finitely) subadditive:
\begin{lemma}\label{lem:subadditivity}
  For any sequence of subsets $E_1,E_2,\ldots$ of $\Omega$,
  \begin{equation*}
    \UpProb
    \left(
      \bigcup_{n=1}^{\infty}
      E_n
    \right)
    \le
    \sum_{n=1}^{\infty}
    \UpProb(E_n).
  \end{equation*}
  In particular,
  a countable union of null sets is null.
\end{lemma}

\section{Main result}
\label{sec:result}

For each $p\in(0,\infty)$,
the \emph{strong} $p$-\emph{variation} of $\omega\in\Omega$ is
\begin{equation*}
  \var_p(\omega)
  :=
  \sup_\kappa
  \sum_{i=1}^n
  \left|
    \omega(t_i)-\omega(t_{i-1})
  \right|^p,
\end{equation*}
where $n$ ranges over all positive integers
and $\kappa$ over all subdivisions $0=t_0<t_1<\cdots<t_n=T$
of the interval $[0,T]$.
It is obvious that there exists a unique number
$\vex(\omega)\in[0,\infty]$,
called the \emph{strong variation exponent} of $\omega$,
such that $\var_p(\omega)$ is finite when $p>\vex(\omega)$
and infinite when $p<\vex(\omega)$;
notice that $\vex(\omega)\notin(0,1)$.

The following is a game-theoretic counterpart
of the well-known property of continuous semimartingales
(Lepingle \cite{lepingle:1976},
Theorem 1 and Proposition 3;
L\'evy \cite{levy:1940} in the case of Brownian motion).
\begin{theorem}\label{thm:main}
  For almost all $\omega\in\Omega$,
  \begin{equation}\label{eq:main}
    \vex(\omega)=2 \text{ or $\omega$ is constant}.
  \end{equation}
\end{theorem}
\noindent
(Alternatively, (\ref{eq:main}) can be expressed as $\vex(\omega)\in\{0,2\}$.)

\section{Proof}
\label{sec:proof}

The more difficult part of this proof
($\vex(\omega)\le2$ a.s.)\ will be modelled on the proof in \cite{bruneau:1979},
which is surprisingly game-theoretic in character.
The proof of the easier part is modelled on \cite{GTP5}.
(Notice, however, that our framework
is very different from those of \cite{bruneau:1979} and \cite{GTP5},
which creates additional difficulties.)
Without loss of generality we impose the restriction $\omega(0)=0$.

\subsection*{Proof that $\vex(\omega)\ge2$ for non-constant $\omega$ a.s.}

We need to show that the event $\vex(\omega)<2 \And \nc(\omega)$ is null,
where $\nc(\omega)$ stands for ``$\omega$ is not constant''.
By Lemma \ref{lem:subadditivity}
it suffices to show that $\vex(\omega)<p \And \nc(\omega)$ is null
for each $p\in(0,2)$.
Fix such a $p$.
It suffices to show that $\var_p(\omega)<\infty \And \nc(\omega)$ is null
and, therefore, it suffices to show that the event
$\var_p(\omega)<C \And \nc(\omega)$ is null for each $C\in(0,\infty)$.
Fix such a $C$.
Finally, it suffices to show that the event
\begin{equation*}
  E_{p,C,A}
  :=
  \left\{
    \omega\in\Omega
    \Biggm|
    \var_p(\omega)<C \And \sup_{t\in[0,T]}\lvert\omega(t)\rvert>A
  \right\}
\end{equation*}
is null for each $A>0$.
Fix such an $A$.

Choose a small number $\delta>0$ such that $A/\delta\in\bbbn$,
and let $\Gamma:=\{k\delta\st k\in\bbbz\}$
be the corresponding grid.
Define a sequence of stopping times $\tau_n$ inductively by
\begin{equation*}
  \tau_{n+1}
  :=
  \inf
  \bigl\{
    t>\tau_n
    \bigm|
    \omega(t)\in\Gamma\setminus\{\omega(\tau_n)\}
  \bigr\},
  \quad
  n=0,1,\ldots,
\end{equation*}
with $\tau_0:=0$
and $\inf\emptyset$ understood to be $\infty$.
Set $T_A:=\inf\{t\st\lvert\omega(t)\rvert=A\}$,
again with $\inf\emptyset:=\infty$,
and
\begin{equation*}
  h_n(\omega)
  :=
  \begin{cases}
    2\omega(\tau_n) & \text{if $\tau_n(\omega)<T\wedge T_A(\omega)$ and $n+1<C/\delta^p$}\\
    0 & \text{otherwise}.
  \end{cases}
\end{equation*}
The elementary capital process
corresponding to the elementary gambling strategy
$G:=(\tau_n,h_n)_{n=1}^{\infty}$
and initial capital $c:=\delta^{2-p}C$
will satisfy
\begin{multline*}
  \omega^2(\tau_{n+1})-\omega^2(\tau_{n})
  =
  2\omega(\tau_{n})
  \left(
    \omega(\tau_{n+1})-\omega(\tau_{n})
  \right)
  +
  \left(
    \omega(\tau_{n+1})-\omega(\tau_{n})
  \right)^2\\
  =
  \K^{G,c}_{\tau_{n+1}}(\omega)-\K^{G,c}_{\tau_n}(\omega)
  +
  \delta^2
\end{multline*}
provided $\tau_{n+1}(\omega)\le T\wedge T_A(\omega)$ and $n+1<C/\delta^p$,
and so satisfy
\begin{equation}\label{eq:cumulative}
  \omega^2(\tau_N)
  =
  \K^{G,c}_{\tau_N}(\omega)
  -
  \K^{G,c}_0
  +
  N\delta^2
  =
  \K^{G,c}_{\tau_N}(\omega)
  -
  \delta^{2-p} C
  +
  \delta^{2-p} N \delta^p
  \le
  \K^{G,c}_{\tau_N}(\omega)
\end{equation}
provided $\tau_N(\omega)\le T\wedge T_A(\omega)$ and $N<C/\delta^p$.
On the event $E_{p,C,A}$ we have $T_A(\omega)<T$
and $N<C/\delta^p$ for the $N$ defined by $\tau_N=T_A$.
Therefore, on this event
\begin{equation*}
  A^2
  =
  \omega^2(T_A)
  \le
  \K^{G,c}_{T_A}(\omega)
  =
  \K^{G,c}_T(\omega).
\end{equation*}
We can see that $\K^{G,c}_t(\omega)$ increases from $\delta^{2-p}C$,
which can be made arbitrarily small by making $\delta$ small,
to $A^2$ over $[0,T]$;
this shows that the event $E_{p,C,A}$ is null.

The only remaining gap in our argument is that
$\K^{G,c}_t$
may become strictly negative
strictly between some $\tau_n<T\wedge T_A$ and $\tau_{n+1}$
with $n+1<C/\delta^p$
(it will be positive at all $\tau_N\in[0,T\wedge T_A]$ with $N<C/\delta^p$,
as can be seen from (\ref{eq:cumulative})).
We can, however, bound $\K^{G,c}_t$ for $\tau_n<t<\tau_{n+1}$
as follows:
\begin{equation*}
  \K^{G,c}_t(\omega)
  =
  \K^{G,c}_{\tau_n}(\omega)
  +
  2\omega(\tau_{n})
  \left(
    \omega(t)-\omega(\tau_{n})
  \right)
  \ge
  2\lvert\omega(\tau_{n})\rvert
  \left(
    -\delta
  \right)
  \ge
  -2A\delta,
\end{equation*}
and so we can make the elementary capital process positive
by adding the negligible amount $2A\delta$ to Sceptic's initial capital.

\subsection*{Proof that $\vex(\omega)\le2$ a.s.}

We need to show that the event $\vex(\omega)>2$ is null,
i.e., that $\vex(\omega)>p$ is null for each $p>2$.
Fix such a $p$.
It suffices to show that $\var_p(\omega)=\infty$ is null,
and therefore,
it suffices to show that event
\begin{equation*}
  E_{p,A}
  :=
  \left\{
    \omega\in\Omega
    \Biggm|
    \var_p(\omega)=\infty \And \sup_{t\in[0,T]}\lvert\omega(t)\rvert<A
  \right\}
\end{equation*}
is null for each $A>0$.
Fix such an $A$.

The rest of the proof follows \cite{bruneau:1979} closely.
Let $M_t(f,(a,b))$ be the number of upcrossings of the open interval $(a,b)$
by a continuous function $f\in\Omega$ during the time interval $[0,t]$,
$t\in[0,T]$.
For each $\delta>0$ we also set
\begin{equation*}
  M_t(f,\delta)
  :=
  \sum_{k\in\bbbz}
  M_t(f,(k\delta,(k+1)\delta).
\end{equation*}
The strong $p$-variation $\var_p(f,[0,t])$ of $f\in\Omega$
over an interval $[0,t]$, $t\le T$, is defined as
\begin{equation*}
  \var_p(f,[0,t])
  :=
  \sup_\kappa
  \sum_{i=1}^n
  \left|
    f(t_i)-f(t_{i-1})
  \right|^p,
\end{equation*}
where $n$ ranges over all positive integers
and $\kappa$ over all subdivisions $0=t_0<t_1<\cdots<t_n=t$
of the interval $[0,t]$
(so that $\var_p(f)=\var_p(f,[0,T])$).
The following key lemma is proved in \cite{bruneau:1979}
(Lemma 1;
in fact, this lemma only requires $p>1$).
\begin{lemma}\label{lem:Bruneau}
  For all $f\in\Omega$, $t>0$, and $q\in[1,p)$,
  \begin{equation*}
    \var_p(f,[0,t])
    \le
    \frac{2^{p+q+1}}{1-2^{q-p}}
    \left(
      2c_{q,\lambda,t}(f)
      +
      1
    \right)
    \lambda^p,
  \end{equation*}
  where
  \begin{equation*}
    \lambda
    \ge
    \sup_{s\in[0,t]}
    \lvert f(s)-f(0) \rvert
  \end{equation*}
  and
  \begin{equation*}
    c_{q,\lambda,t}(f)
    :=
    \sup_{k\in\bbbn}
    2^{-kq}
    M_t(f,\lambda 2^{-k}).
  \end{equation*}
\end{lemma}
Another key ingredient of the proof is the following game-theoretic version
of Doob's upcrossings inequality:
\begin{lemma}\label{lem:Doob}
  Let $c<a<b$ be real numbers.
  For each elementary capital process $S\ge c$
  there exists a positive elementary capital process $S^*$
  that starts from $S^*_0=a-c$ and satisfies,
  for all $t\in[0,T]$ and $\omega\in\Omega$,
  \begin{equation*}
    S^*_t(\omega)
    \ge
    (b-a)M_t(S(\omega),(a,b)),
  \end{equation*}
  where $S(\omega)$ stands for the sample path $t\mapsto S_t(\omega)$.
\end{lemma}
\begin{proof}
  The following standard argument is easy to formalize.
  Let $G$ be an elementary gambling strategy leading to $S$
  (when started with initial capital $S_0$).
  An elementary gambling strategy $G^*$ leading to $S^*$
  (with initial capital $a-c$)
  can be defined as follows.
  When $S$ first hits $a$,
  $G^*$ starts mimicking $G$ until $S$ hits $b$,
  at which point $G^*$ chooses portfolio $0$;
  after $S$ hits $a$,
  $G^*$ mimics $G$ until $S$ hits $b$,
  at which point $G^*$ chooses portfolio $0$;
  etc.
  Since $S\ge c$,
  $S^*$ will be positive.
\end{proof}

Now we are ready to finish the proof of the theorem.
Let $T_A:=\inf\{t\st\omega(t)=A\}$ be the hitting time for $A$
(with $T_A:=T$ if $A$ is not hit).
By Lemma \ref{lem:Doob},
for each $k\in\bbbn$ and each $i\in\{-2^k+1,\ldots,2^k\}$
there exists a positive elementary capital process $S^{k,i}$
that starts from $A+(i-1)A2^{-k}$ and satisfies
\begin{equation*}
  S^{k,i}_{T_A}
  \ge
  A2^{-k}
  M_{T_A}
  \left(
    \omega,
    \left(
      (i-1)A2^{-k},iA2^{-k}
    \right)
  \right).
\end{equation*}
Summing $2^{-kq}S^{k,i}/A2^{-k}$ over $i\in\{-2^k+1,\ldots,2^k\}$,
we obtain a positive elementary capital process $S^k$ such that
\begin{equation*}
  S_0^k
  =
  2^{-kq}
  \sum_{i=-2^k+1}^{2^k}
  \frac{A+(i-1)A2^{-k}}{A2^{-k}}
  \le
  2^{-kq}
  2^{2k+1}
\end{equation*}
and
\begin{equation*}
  S_{T_A}^k
  \ge
  2^{-kq}
  M_{T_A}(\omega,A2^{-k}).
\end{equation*}
Next, assuming $q\in(2,p)$ and summing over $k\in\bbbn$,
we obtain a positive capital process $S$ such that
\begin{equation*}
  S_0
  =
  \sum_{k=1}^{\infty}
  2^{-kq}
  2^{2k+1}
  =
  \frac{2^{3-q}}{1-2^{2-q}}
  \quad\text{and}\quad
  S_{T_A} \ge c_{q,A,T_A}(\omega).
\end{equation*}
On the event $E_{p,A}$ we have $T_A=T$
and so, by Lemma \ref{lem:Bruneau},
$c_{q,A,T_A}(\omega)=\infty$.
This shows that $S_T=\infty$ on $E_{p,A}$
and completes the proof.

\section{Conclusion}
\label{sec:conclusion}

Theorem \ref{thm:main} says that, almost surely,
\begin{equation*}
  \var_p(\omega)
  \begin{cases}
    <\infty & \text{if $p>2$}\\
    =\infty & \text{if $p<2$ and $\omega$ is not constant}.
  \end{cases}
\end{equation*}
The situation for $p=2$ remains unclear.
It would be very interesting to find the upper probability
of the event
$\{\var_2(\omega)<\infty\text{ and $\omega$ is not constant}\}$.
(L\'evy's \cite{levy:1940} result
shows that this event is null
when $\omega$ is the sample path of Brownian motion,
while Lepingle \cite{lepingle:1976} shows this
for continuous, and some other, semimartingales.)

\subsection*{Acknowledgments}

This work was partially supported by EPSRC (grant EP/F002998/1),
MRC (grant G0301107),
and the Cyprus Research Promotion Foundation.


\begin{thebibliography}{10}

\bibitem{bruneau:1979}
Michel Bruneau.
\newblock Sur la $p$-variation des surmartingales.
\newblock {\em S\'eminaire de probabilit\'es de Strasbourg}, 13:227--232, 1979.
\newblock Available free of change at \texttt{http://www.numdam.org}.

\bibitem{horikoshi/takemura:2008}
Yasunori Horikoshi and Akimichi Takemura.
\newblock Implications of contrarian and one-sided strategies for the fair-coin
  game.
\newblock \emph{Stochastic Processes and their Applications}, to appear,
  doi:10.1016/j.spa.2007.11.007.

\bibitem{kumon/takemura:2008}
Masayuki Kumon and Akimichi Takemura.
\newblock On a simple strategy weakly forcing the strong law of large numbers
  in the bounded forecasting game.
\newblock \emph{Annals of the Institute of Statistical Mathematics}, to appear.

\bibitem{kumon/etal:2007}
Masayuki Kumon, Akimichi Takemura, and Kei Takeuchi.
\newblock Game-theoretic versions of strong law of large numbers for unbounded
  variables.
\newblock {\em Stochastics}, 79:449--468, 2007.

\bibitem{lepingle:1976}
Dominique Lepingle.
\newblock La variation d'ordre $p$ des semi-martingales.
\newblock {\em Zeit\-schrift f\"ur Wahrscheinlichkeitstheorie und verwandte
  Gebiete}, 36:295--316, 1976.

\bibitem{levy:1940}
Paul L\'evy.
\newblock Le mouvement brownien plan.
\newblock {\em American Journal of Mathematics}, 62:487--550, 1940.

\bibitem{shafer/vovk:2001}
Glenn Shafer and \Vladimir{} Vovk.
\newblock {\em Probability and Finance: It's Only a Game!}
\newblock Wiley, New York, 2001.

\bibitem{takeuchi:2004latin}
Kei Takeuchi.
\newblock {\em Kake no suuri to kinyu kogaku (Mathematics of Betting and
  Financial Engineering, in Japanese)}.
\newblock Saiensusha, Tokyo, 2004.

\bibitem{takeuchi/etal:2007}
Kei Takeuchi, Masayuki Kumon, and Akimichi Takemura.
\newblock A new formulation of asset trading games in continuous time with
  essential forcing of variation exponent.
\newblock Technical Report \texttt{arXiv:0708.0275v1} [math.PR],
  \texttt{arXiv.org} e-Print archive, August 2007.

\bibitem{vovk:arXiv0712.1275}
\Vladimir{} Vovk.
\newblock Continuous-time trading and emergence of randomness, {I}.
\newblock Technical Report \texttt{arXiv:0712.1275} [math.PR],
  \texttt{arXiv.org} e-Print archive, December 2007.

\bibitem{GTP5}
\Vladimir{} Vovk and Glenn Shafer.
\newblock A game-theoretic explanation of the $\sqrt{dt}$ effect. {T}he
  {G}ame-{T}heoretic {P}robability and {F}inance project,
  \texttt{http://prob\linebreak[0]a\linebreak[0]bil\linebreak[0]i\linebreak[0]%
ty\linebreak[0]and\linebreak[0]fi\linebreak[0]nance.com}, {W}orking {P}aper 5,
  January 2003.

\end{thebibliography}
\end{document}